\documentclass[mathleft]{an}
\usepackage{graphicx}
\usepackage{times}
\begin{document}

\Pagespan{1}{}
\Yearpublication{2010}%
\Yearsubmission{2009}%
\Month{07}%
\Volume{}%
\Issue{}%
\DOI{}

\title{Observational Evidence for Expansion in the SSS spectra of Novae}

\author{J.-U. Ness}
\titlerunning{Systematic line shifts of X-ray absorption lines in SSS spectra}
\authorrunning{J.-U. Ness}
\institute{European Space Astronomy Centre,
P.O. Box 78, 28691 Villanueva de la Ca\~nada, Madrid, Spain; juness@sciops.esa.int}

\received{\today}
\accepted{11 Nov 2005}
\publonline{later}


\abstract{
 For several novae, a bright X-ray source with a spectrum resembling the
class of Super Soft X-ray Sources (SSS) has
been observed a few weeks to months after outburst.
Novae are powered by explosive nuclear burning on the surface
of a white dwarf, and enough energy is produced to power a radiatively
driven wind. Owing to the evolution of the opacity of the ejecta, the
observable spectrum gradually shifts from optical to soft X-rays (SSS
phase). It has sometimes been assumed that at the beginning of the SSS
phase no more mass loss occurs. However, high-resolution X-ray spectra
of some novae have shown highly blue-shifted absorption lines, indicating
a significant expansion. In this paper, I show that all novae that have
been observed with X-ray gratings during their SSS phase show significant
blue shifts. I argue that all models that attempt to explain the X-ray
bright SSS phase have to accommodate the continued expansion of the ejecta.
}

\maketitle

\vspace{-.3cm}
\section{Introduction}
\vspace{-.3cm}

\begin{figure}[!ht]
\resizebox{\hsize}{!}{\includegraphics{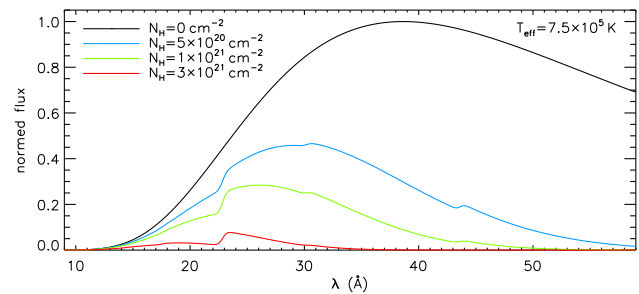}}
\caption{\label{bbsoft}A blackbody model with effective temperature
$T_{\rm eff}=7.5\times10^5$\,K and four different predicted spectra
assuming different values of interstellar neutral hydrogen column
density (absorption model {\tt bamabs} from Kashyap \& Drake 2000).
}
\end{figure}

 An X-ray source is classified as a Super Soft X-ray Source (SSS) if the
X-ray spectrum resembles a blackbody continuum spectrum (\cite{planck01}),
yielding an effective blackbody temperature of less than $\sim 100$\,eV
(several $10^5$\,K, see, e.g., {Kahabka} \& {van den Heuvel} 1997). The Wien
tail of such a spectrum is rarely above 1\,keV, while the Rayleigh-Jeans
tail can not be observed, because the soft emission is absorbed by the
interstellar medium. This is illustrated in Fig.~\ref{bbsoft}, where a
blackbody model is shown with no absorption (black) and after correction
for interstellar absorption assuming three different values of neutral
hydrogen column density, $N_{\rm H}$ (colour-coded lines). The absorption
model as implemented in the PintofAle program package, described by
\cite{pintofale} has been used to compute the corrected spectra.
One can see that observations of SSSs sample only a small fraction
of the source emission, and the shape of the soft part of the spectrum
is dominated by the ISM transmission function.
This is one of the main challenges to observe SSSs in our own
galaxy, which explains why many SSSs, including the two prototypes
Cal\,83 and Cal\,87, are in the LMC, with less galactic foreground
absorption. The downside
of observing extragalactic SSSs is the larger distance, which dilutes the
brightness.

 Meanwhile, Classical Novae have been observed to undergo a
phase of SSS emission when their X-ray spectra resemble those of
the persistent SSSs. Since novae during the SSS phase are much brighter
than  persistent SSSs, they can be observed at larger distance and with
better exposed spectra, yielding important clues about the properties of
the emitting plasma.

\begin{figure*}[!t]
\resizebox{\hsize}{!}{\includegraphics{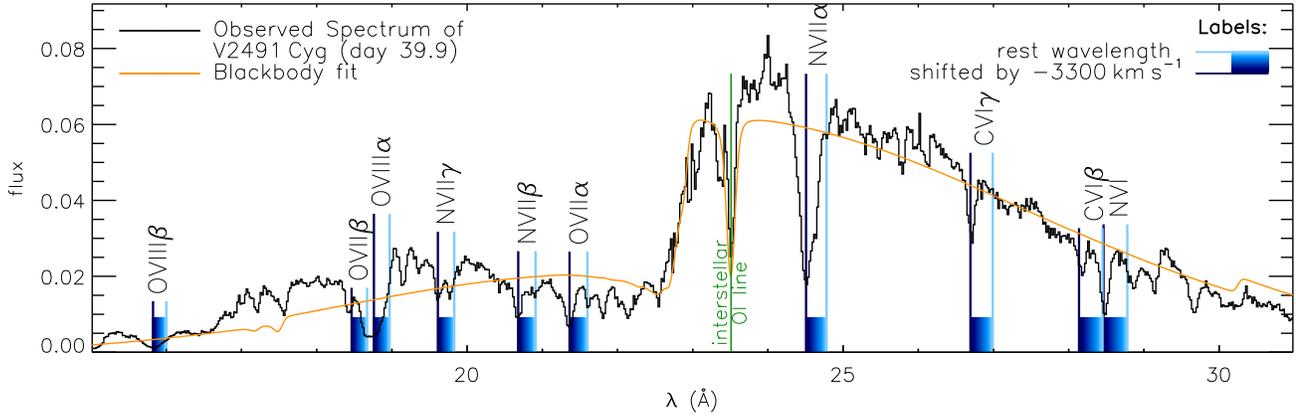}}
\caption{\label{allspec}XMM-Newton RGS spectrum of V2491\,Cyg. Flux units
are photons\,cm$^{-2}$\,s$^{-1}$\,\AA$^{-1}$. The continuum resembles
a blackbody (orange curve), but in addition, deep absorption lines can
be identified. The blackbody curve is corrected for interstellar
absorption using the ISM absorption model by Wilms et al. (2000).
 The blue-shaded boxes indicate a range of blue shifts ranging
from rest wavelength (light) up to a velocity of
3300km s$^{-1}$ (dark).
}
\end{figure*}

 Novae occur in close binary systems with an accreting white dwarf.
After a history of quiet accretion, the point is reached that the
accreted hydrogen-rich material can ignite in a thermonuclear
runaway. Radiation pressure leads to the ejection of envelope material,
forming an optically thick shell. The high-energy radiation produced
by nuclear burning has to travel through this material via
radiation transport. The resultant spectrum is thus an atmospheric
spectrum that originates from the photosphere, and it consists of a
blackbody-like continuum with absorption lines. As the mass loss rate
decreases, the density of the outer layers decreases, and they become
optically thin, thus allowing observations deeper into the outflow.
The observed effective temperature thus depends on the radius of
the photosphere that determines which regions are observable. As
time progresses, successively hotter layers are exposed to
observations, and the peak of the spectral energy distribution
eventually reaches the X-ray regime.

 As a first approach to interpret observed SSS spectra, blackbody
models are routinely fitted, mainly for reasons that they are
easy to do, and that they are quickly reproducible so the blackbody
temperature can serve as a spectral classification criterion.
However, the implied assumption of thermal equilibrium is not satisfied,
and complex absorption processes are completely ignored.
One can thus hardly consider a blackbody fit a suitable model, 
however, the low spectral resolution of less than 100 and low
signal to noise of many available X-ray spectra do not allow us
to find sensible constraints to more sophisticated models.

 Krautter et al. (1996) derived a bolometric lumi\-no\-sity from blackbody
fits that is up to 100 times the Eddington luminosity and argued
that this is unphysically high. Atmosphere models assuming
local thermodynamic equilibrium (LTE) have been applied by \cite{balm98},
yielding a lower luminosity. However, the spectral resolution of
the ROSAT observation of only $E/\Delta E=2.3$ at 0.93\,keV is
not sufficient to constrain any atmosphere model. For comparison,
the spectral resolution of XMM-Newton/RGS is $E/\Delta E=240$ at
1\,keV. It has further not been discussed whether the assumption
of LTE is valid. For example, {Hartmann} \& {Heise} (1997) showed that
LTE and non-LTE models can produce very different model spectra
where the temperature of the non-LTE model can be closer the
blackbody temperature than that derived from the LTE model.
\cite{rauch05a} computed a non-LTE model for an X-ray grating spectrum of
the nova V4743\,Sgr during the SSS phase. Reasonable agreement
with the grating spectrum was found, and temperature and
elemental abundances were determined. However, the atmosphere
model was that of a white dwarf with the plane-parallel
approximation, and in order to
match blue shifts of the absorption lines that had previously
been reported by {Ness} {et~al.} (2003), they assumed a high radial
velocity of the entire system of 2400km s$^{-1}$. \cite{petz05}
developed a non-LTE model that accounts for expansion of the
ejecta, thus modelling the line shifts in a self-consistent way.
The radiation transport processes can be much different from those
in a static, plane-parallel white dwarf atmosphere. The models by
{Petz} {et~al.} (2005) are currently being improved by Daan van Rossum.
The purpose of this paper is to gather all observational
evidence that justifies the need for such models.

\vspace{-.6cm}
\section{Observations} 
\vspace{-.3cm}

 For this work grating observations of the galactic novae
V4743 Sgr, RS Oph, and V2491 Cyg are used which have
been presented by {Ness} {et~al.} (2003), {Ness} {et~al.} (2007),
{Ness} {et~al.} (2009), and Ness et al. in preparation, respectively.
In  Fig.~\ref{allspec}, the XMM-Newton RGS spectrum of
V2491\,Cyg, taken on day 39.9 is shown as an example. In
Table~\ref{obslog}, the targets and observations are summarised,
giving time of outburst and distance, instrument used for
observations, time of the observations (in days after
outburst), and exposure time and, in the last three columns,
best-fit blackbody parameters. The correction for interstellar
absorption was done with the model by \cite{wilms00}, that includes
the O\,{\sc i} 1s-2p line at 23.51\,\AA. In Fig.~\ref{allspec}, the best-fit
blackbody curve to the spectrum of V2491\,Cyg is shown, and
it can clearly be seen that many absorption features are
not fitted with a blackbody model.
Note also from Table~\ref{obslog} that the derived bolometric
luminosity is in all cases higher than the Eddington luminosity
of a 1-M$_\odot$ white dwarf of $\log L_{\rm bol}\approx 38$.

\begin{table*}[!ht]
\begin{center}
\caption{\label{obslog}Target characteristics and observations}
\begin{tabular}{lllllllll}
\hline
Nova & day of Outburst$^a$ & $d^b$ & Instrument & day$^c$ & $\Delta t^d$ & $T_{\rm bb}^e$ & $N_{\rm H}^f$ & $\log L_{\rm bol}^g$\\
V4743\,Sgr & 2002 Sept. 20.4 & 3.9 & Chandra/LETGS & 180.4 & 15 & 3.9 & 2.1 & 40.21\\
&&& Chandra/LETGS & 371.0 & 12 & 4.6 & 1.8 & 39.25 \\
RS\,Oph & 2006 Feb 12.8 & 1.6 & Chandra/LETGS & 39.7 & 10 & 6.2 & 5.1 & 38.43 \\
&&& Chandra/LETGS & 66.9 & 6.5 & 5.0 & 6.9 & 39.86\\
V2491\,Cyg & 2008 Apr 10.7 & 10.5 & XMM-Newton/RGS & 39.9 & 39 & 5.2 & 5.0 & 40.02 \\
&&& XMM-Newton/RGS & 49.7 & 30 & 5.7 & 4.7 & 38.94\\
\hline
\end{tabular}
\end{center}

$^a$Time of optical discovery, treated as start time of the outburst\hfill
$^b$Distance in kpc\\
$^c$Time of observation in units days after outburst\hfill
$^d$Exposure time in $10^3$\,s\\
$^e$Blackbody temperature in $10^5$K\hfill
$^f$Neutral hydrogen column density in $10^{21}$\,cm$^{-2}$\\
$^g$Logarithmic bolometric luminosity in erg\,s$^{-1}$
\end{table*}

 In the XMM-Newton RGS spectrum of V2491\,Cyg that is shown in
Fig.~\ref{allspec}, one can clearly see that strong absorption lines
are present. Most lines originate from H-like and He-like lines of
C, N, and O. Some of these lines are marked in Fig.~\ref{allspec}.
The shadings, ranging from light to dark blue, indicate the range
between the rest wavelengths (light) to the expected wavelengths
assuming a Doppler velocity of  $-3300$km s$^{-1}$ (dark).
It can clearly be seen that for the indicated lines, deep
absorption lines occur shortwards of the rest wavelengths but
in agreement with the blue-shifted wavelengths. For the C\,{\sc vi}
Ly$\beta$ (1s-3p) line at 28.45\,\AA, a deep line occurs at the
rest wavelength,
but the identification as blue-shifted N\,{\sc vi} Ly$\alpha$ (1s-2p) 
appears more likely in light of the arguable blue shifts of
all the other lines and the given depth of the N\,{\sc vii}
Ly$\alpha$ line around 24.8\,\AA.

\begin{figure}[!b]
\resizebox{\hsize}{!}{\includegraphics{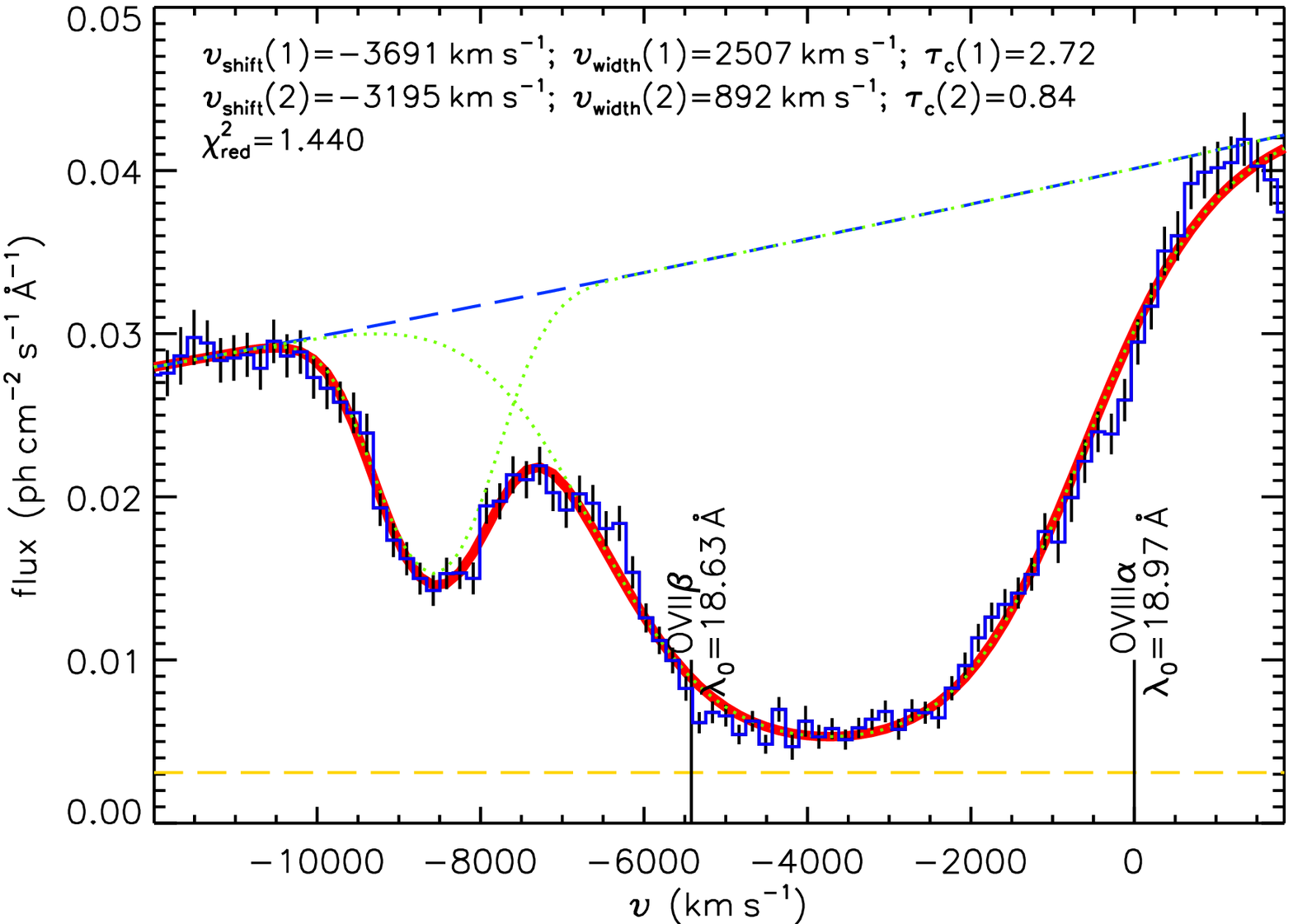}}

\resizebox{\hsize}{!}{\includegraphics{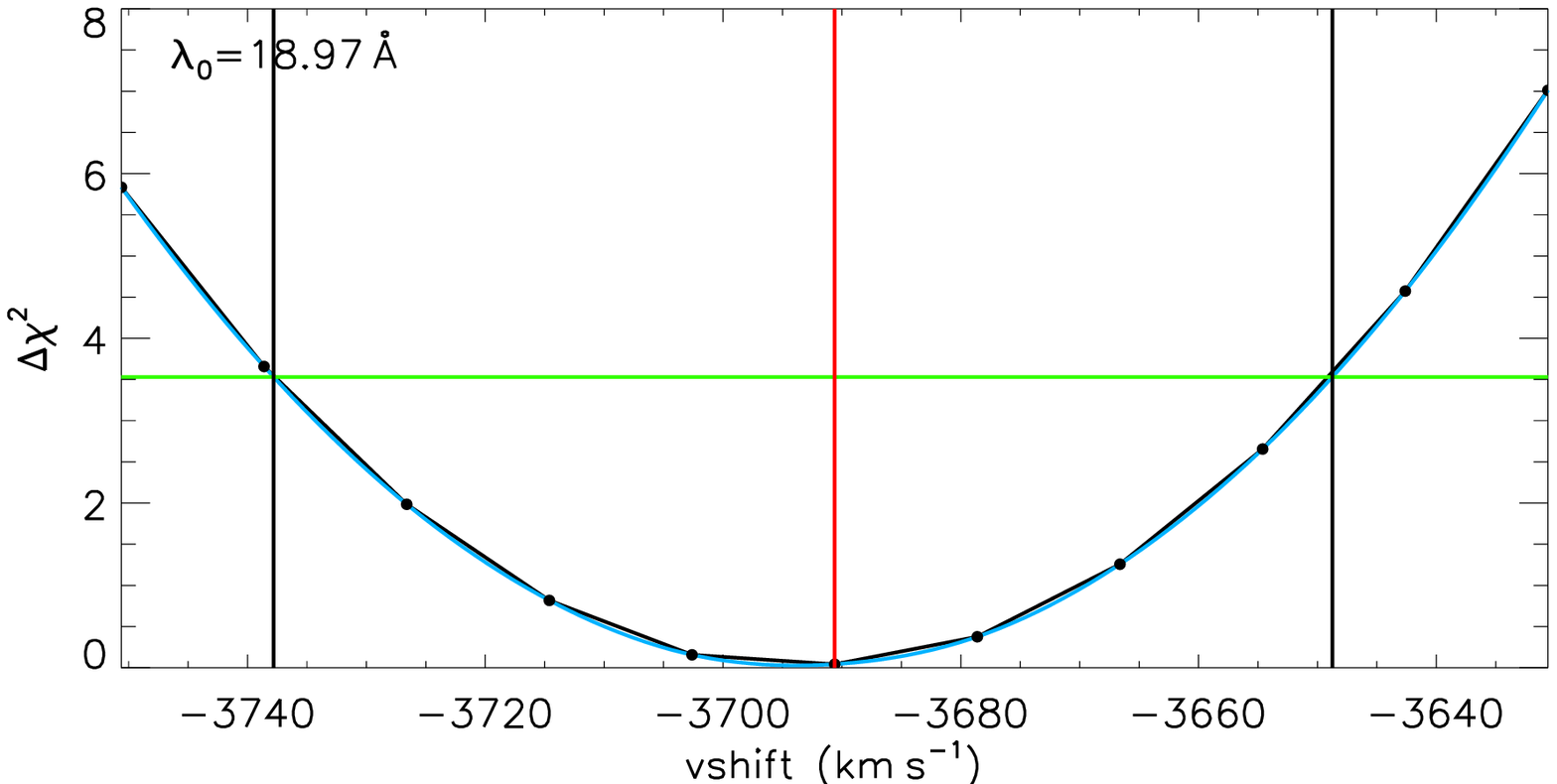}}
\caption{\label{lfit}As an illustration of the line fitting method, the
fit to the O\,{\sc viii} 1s-2p line ($\lambda_0=18.97$\,\AA) and the
O\,{\sc vii} 1s-3p line ($\lambda_0=18.63$\,\AA) for V2491\,Cyg is
shown. In the top panel, the spectrum with the best-fit curve is
shown. The x-axis is Doppler velocity based on the rest wavelength
of O\,{\sc viii}. The positions of the rest wavelengths are marked.
The O\,{\sc viii} line is clearly saturated, but the emission
level does not drop to zero. In the bottom panel the value
$\Delta \chi^2=\chi^2-\chi^2_{\rm best}$ as a function of
assumed line shift is shown. The best fit is marked with the vertical
red line, and the 68.3\% uncertainty range is marked with the
black vertical lines.
}
\end{figure}

 In order to determine the exact line shifts and line widths with
uncertainties, I have selected the strongest and most isolated
(non-blended) lines in the spectra of V4743\,Sgr, V2491\,Cyg,
and RS\,Oph and fitted the narrow spectral regions around
each line with the curve
\begin{equation}
   C(\lambda)\times {\rm e}^{-\tau(\lambda)} \times T(\lambda)\,,
\end{equation}
 where $C(\lambda)$, $\tau(\lambda)$, and $T(\lambda)$ are the continuum,
optical dep\-th, and interstellar transmission, as function of
wavelength, $\lambda$, respectively. For the continuum, a blackbody
curve with adjustable temperature, and normalisation was used. The
transmission curve was calculated using the tool {\tt bamabs}
by \cite{pintofale} with the free parameter $N_{\rm H}$.
The optical depth is defined as
\begin{equation}
   \tau(\lambda)=\tau_{\rm c}{\rm e}^{(\lambda-\lambda_{\rm c})^2/\sigma^2}\,,
\end{equation}
with $\tau_{\rm c}$ the optical depth at line centre,
$\lambda_{\rm c}$ the central wavelength, and $\sigma$ the Gaussian
line width.

 In Fig.~\ref{lfit}, I show, as an example, the observation of
V2491\,Cyg on day 39.7 and the best fit to the
the O\,{\sc viii} (1s-2p) and O\,{\sc vii} (1s-3p)
lines at $\lambda_0=18.97$\,\AA\ and $\lambda_0=18.63$\,\AA,
respectively. On the x-axis, a velocity scale is sho\-wn,
assuming that 18.97\,\AA\ is the rest wavelength. The best-fit velocities
derived from the line-shifts and line-widths are given in the
legend, together the value of reduced $\chi^2$. Note that for the
purposes of determining line shifts and line widths, the continuum
model is not important, and the additional free parameters can be
regarded as 'uninteresting'; see \cite{avni76}.

 Even without the fit, one can already see that the O\,{\sc viii} line
is saturated, however, the emission level in the line centre never
drops to zero. Therefore, significant continuum emission that
arises from above the O\,{\sc viii}-absorbing material must be
present, and this possibility has been accounted for in the fit
by adding a constant flux in all spectral bins. This emission level
was a free parameter, and the best-fit emission level of this
additional continuum emission is indicated by the dashed yellow
line in Fig.~\ref{lfit}.

 With a value of reduced $\chi^2_{\rm red}=1.4$, the fit is acceptable,
however, one can see that the profile of the O\,{\sc viii} line
is not smooth, indicating that the different absorbing layers are not
uniform in column density. While the line shift can be determined
relatively accurately, the line widths may be affected by larger
systematic uncertainties arising from the irregular shape of the
profile and also by blends with weaker lines. In the bottom panel
of Fig.~\ref{lfit}, I illustrate how the uncertainties for the
line shift of the O\,{\sc viii} line has been derived. The best-fit
line shift is marked by the vertical red line in the centre, and
the parabolic curve is the difference of
$\Delta \chi^2(v)=\chi^2(v)-\chi^2_{\rm best}$ as a function of
velocity $v$. The black bullets mark grid points at which a fixed
line shift was assumed while all other parameters were refitted.
With three free parameters (line shift, line width, and line
depth)\footnote{All other parameters are 'uninteresting'; see Avni
(1976)}, the value of $\chi^2$ does not increase
by more than 3.53 (horizontal green line in the bottom panel of
Fig.~\ref{lfit}) if the line shift is within the 1-$\sigma$
(=68.3\%) uncertainty range. This range is indicated by the vertical
black line to the right and left of the best-fit value.

\vspace{-.6cm}
\section{Results and Discussion}
\vspace{-.3cm}

\begin{figure}[!t]
\resizebox{\hsize}{!}{\includegraphics{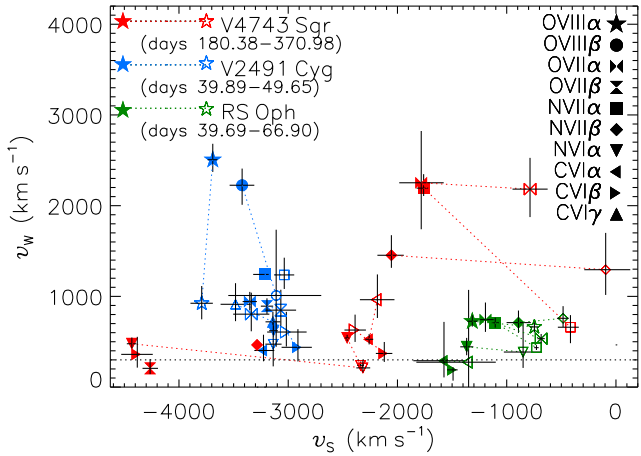}}
\caption{\label{shifts}Best-fit line shifts (abscissa) and
line widths (ordinate) for V4743\,Sgr (red), V2491\,Cyg (blue),
and RS\,Oph (green). In the upper right legend, the plot symbols
are explained, where the Greek letters stand for the transitions
1s-2p ($\alpha$), 1s-3p ($\beta$), and 1s-4p ($\gamma$).
The evolution of velocities from two measurements
of the same line are indicated by connecting symbols with
dotted lines.
}
\end{figure}

 The results are presented in Fig.~\ref{shifts}, where line shifts
versus line widths are shown. The main result is that {\em all} lines
are blue shifted, and that the observed line shifts change with time.
One can also see that different velocity systems can be identified from
the X-ray spectra. While for RS\,Oph, all lines are consistently blue
shifted by $\sim 1200-1600$km s$^{-1}$, the observed
velocities in V2491\,Cyg are much higher, with more than
3000km s$^{-1}$. For V4743 Sgr, the situation is more
complicated, and more than one velocity system can be identified
in some lines. Blue shifts of more than 4000km s$^{-1}$ can
easily be mixed up with other lines, however, the presence of such
a fast component in three different lines suggests that this
component is indeed present. With the presence of different velocity
components, the interpretation of the line widths has to be
approached with more care and will be discussed in more detail in
upcoming papers.

 The detection of blue shifts can either be interpreted as the
radial velocity of the system or as the expansion velocity of the
ejecta. The former interpretation has implicitly been assumed by, e.g.,
\cite{rauch05a} in order to bring static atmosphere
models in agreement with the observations of V4743\,Sgr. However,
a radial velocity of 2400km s$^{-1}$ seems unreasonably
high. Especially the fact that the observed blue shifts change
by several hundreds km\,s$^{-1}$ within only a few weeks is
reason enough to confidently rule out this possibility. The
only possible conclusion from the line shifts is that the
ejecta are still expanding, and that mass loss is thus still
continuing.

 This conclusion has important implications for several model
approaches. For example, {Hachisu} \& {Kato} (2009) assume for their
light curve models that in V2491\,Cyg, the wind has stopped on
or before day 40 after outburst (see their figure 1). Yet, from
Fig.~\ref{shifts} (blue open symbols) one can clearly see that
on day 49.7, significant expansion was still detected. While
for V2491\,Cyg, the change in blue shift from day 39.9 to 49.7
is not significant enough to clearly rule out a radial velocity
of the system of $\sim 3000$km s$^{-1}$, the situation is
clearly different for RS\,Oph, where the radial velocity can
not be higher than $\sim 800$km s$^{-1}$ (see open green
symbols in Fig.~\ref{shifts}), and at least on day 39.7,
the wind had not stopped yet, opposed to the assumptions
formulated by \cite{hachisu_rsoph}.

 Furthermore, most atmosphere models do not account for the
expansion. While it is beyond the scope of this paper to test
the importance of the expansion, significant effects can not
easily be ruled out. It is easy to imagine that radiation
transport works differently in a moving atmosphere, and it
is thus mandatory that models that account for the expansion
are tested. These tests are presented in the same volume by
\cite{vrossum}.

\acknowledgements
I thank Mike Bode for comments on the draft




\end{document}